\documentclass[aps,pra,amsmath,amssymb,11pt,final,tightenlines,twoside,twocolumn,nofloats,nofootinbib,superscriptaddress,
showkeys,showkeywords]{revtex4-2}

\usepackage[T2A]{fontenc}
\usepackage[utf8]{inputenc}
\usepackage[russian,english]{babel}
\usepackage{graphicx}

\usepackage{xcolor}

\newcommand{\WI}[2]{#1_{\mathrm{#2}}}

\begin{document}




\title{Rotating Polytropic Twins and the Fragmentation Problem}

\author{\bf\copyright~2025 г. \firstname{V.}~\surname{Valiulin}}
\affiliation{Vereshchagin Institute of High Pressure Physics, Russian Academy of Sciences, Troitsk, 142190 Russia}
\affiliation{National Research Center Kurchatov Institute, Moscow, 123098 Russia}

\author{\bf\firstname{K.}~\surname{Manukovskiy}}
\affiliation{National Research Center Kurchatov Institute, Moscow, 123098 Russia}

\author{\bf\firstname{T.}~\surname{Razinkova}}
\affiliation{National Research Center Kurchatov Institute, Moscow, 123098 Russia}

\author{\bf\firstname{A.}~\surname{Yudin}}
\email{yudin@itep.ru}
\affiliation{National Research Center Kurchatov Institute, Moscow, 123098 Russia}

\begin{abstract}
\vspace{3mm}
\received{???}
\revised{???}
\accepted{???}
\vspace{3mm}

Polytropic $n=1$ stellar configurations with strong differential rotation have been considered. The
so-called twin configurations: stars having almost identical integral characteristics, but clearly structurally
different have been found. Hydrodynamic 3D modeling has been performed for two such structures and has
shown that these configurations also differ in the dynamic aspect: one of them remains practically stationary,
while the other turns out to be unstable with respect to the development of the bar mode and fragmentation.
The gravitational wave signal accompanying the fragmentation process has been calculated.

\end{abstract}

\maketitle

\section{INTRODUCTION}
A study of stellar models with a polytropic equation of state
$P=K\rho^\gamma$, where $\gamma\equiv 1{+}1/n$, and $n$ is the
polytropic index, has an extensive history \cite{Horedt2004} and is
actively continued at the present time. A polytropic
star in many cases is a starting point for estimating
parameters and developing more accurate models of
real stars. However, in a number of cases, the polytrope
turns out to be more than just a toy model for
a star: for example, white dwarfs with non-relativistic
electrons are perfectly described by the polytrope of
$n=3/2$, while dwarfs with relativistic ones are
described by polytrope of $n=3$, which produces the
Chandrasekhar limit. The parameters of neutron stars
(NSs), despite the importance of the effects of general
relativity (GR) in them, in some aspects can be well
approximated by Newtonian polytropes with $n=1$ \cite{Hensel2007}.

The path that led us to this research was sufficiently
winding. We were initially interested in the question of
whether rotational effects could significantly affect the
parameters of low-mass NSs \cite{Yudin2019}. This question is
important in the context of the so-called stripping
model for short gamma-bursts (see \cite{Blinnikov2021} and references
therein). We then decided to explore the issue of influence
of differential rotation on the structure of the star
(in \cite{Yudin2019}, the rotation was assumed to be rigid-body).
This time, for maximal generality, we used not the
realistic state equation of the NS \cite{Haensel2004}, as in \cite{Yudin2019}, but the
polytropic equation with $\gamma=2$. As a result, we discovered
the existence of a large number of differentially
rotating stellar configurations of various types: tori,
washers, ``matryoshka dolls'', etc. \cite{Razinkova2024}. The key
moment that led to the appearance of this publication
was the discovery that among the configurations
found in \cite{Razinkova2024}, there are so-called ``doubles'' or twins: stars that,
while possessing practically identical integral characteristics
(such as angular momentum, total energy,
etc.), at the same rotation law, are nevertheless clearly
structurally different. The first part of this paper is
devoted to their description.

The second part arose from the desire to verify the
reality of the existence of doubles in some independent
way. Specifically, there was needed a way to
check the correctness of the calculation of these axial–
symmetrical stationary rotating stars obtained by us
using the \textsf{ROTAT} code~\cite{AksBlinn}. For verification, we used
the well-known astrophysical code \textsf{FLASH}~\cite{FLASH}. Our
3D calculations, the details of which are given below,
made it possible both to verify the hydrodynamic stability
of our configurations and to study the problem of
their secular stability with respect to fragmentation.

The third and final part of this paper is devoted to
the question of whether gravitation-wave losses
interfere with the fragmentation process. To do this,
we scaled our polytropic configurations to parameters
typical of neutron stars (mass is the order of the solar
mass, and radius is the order of 10~km) and calculated
off-line the total angular momentum carried away by
gravitational waves during fragmentation.

In conclusion, we discuss the obtained results, the
identified problems (in particular, with the fragmentation
parameter $\tau$) and discuss plans for further
activities.

\section{MAIN INPUT PARAMETERS}
\label{Base}
We briefly describe the input parameters and the
problem statement (for more details, see \cite{Razinkova2024}). We are
operating with a polytropic equation of state $P=K\rho^{\gamma}$,
where $P$ is the pressure, $\rho$ is the density,
and $K=\mathrm{const}$. We remind that, from here, we
accept $\gamma=\mathrm{const}\equiv 1{+}1/n$, where $n$ is the polytropic
index~\cite{Horedt2004}.

We calculate the structure of a star in stationary
rotation using the \textsf{ROTAT} code~\cite{AksBlinn}. Given the equation
of state and the law of rotation (see below), it
makes it possible to find the axial–symmetric solution
for the structure of the star by setting the value of the
configuration compression parameter to be
$\theta\equiv\WI{R}{p}/\WI{R}{e}$, where $\WI{R}{p}$ is the its polar radius, and $\WI{R}{e}$ is
the equatorial radius. A value of $\theta=1$ obviously
responds a spherically–symmetrical star.

We use the following frequently applied law of rotation
(for further arguments in its favor, see~\cite{Razinkova2024}):
\begin{equation}
\omega(\xi)=\frac{\omega_0}{1+(\sigma{-}1)\frac{\xi^2}{\WI{R}{e}^2}}.\label{rotlaw}
\end{equation}
Here $\omega$ is the angular velocity of the star’s rotation,
while $\xi$ is the cylindrical radius (the distance from a
given point to the axis of rotation). According to the
Poincare theorem (see, for example,~\cite{Tassul,Krat}), during
stationary rotation the angular velocity $\omega$ must be a
function of $\xi$ only.

Next, the quantities $\omega_0$ and $\sigma$ in formula (\ref{rotlaw}) are
constants. Parameter $\omega_0$ is automatically determined
by the \textsf{ROTAT} code once the configuration compression $\theta$
is set. Parameter $\sigma$ is set by the user and determines
the degree of differential rotation: it shows how
much faster the center rotates relative to the last point
on the star’s equator at $\xi=\WI{R}{e}$. A value of $\sigma=1$ obviously
corresponds to rigid-body rotation. In this study,
we consider the results only for $\sigma=2$: as we have
shown in our study \cite{Razinkova2024}, this case is the most interesting.

\section{DIMENSIONLESS UNITS}
\label{Sec:Dimensionless}
To proceed further, we need to define units for various
important physical quantities. Their introduction
allows scaling and comparison of different polytropic
stellar configurations. As is well known (see, for example,
\cite{Horedt2004}), for a polytrope of $n=1$, radius $r_0$ is an eigenvalue
of the problem and does not depend on the central
density $\WI{\rho}{c}$:
\begin{equation}
r_0=\sqrt{\frac{K}{2\pi G}}, \label{r0}
\end{equation}
where $G$ is the gravitational constant.

Then, the unit of total mass of the star is
\begin{equation}
M_0=4\pi r_0^3\WI{\rho}{c}.\label{M0}
\end{equation}
As can be seen, the mass linearly scales with the density
at the center.

By unit angular frequency $\omega_0$, it is
natural to accept a value inversely proportional to the
hydrodynamic time of the star, $\WI{t}{h}$:
\begin{equation}
\omega_0=1/\WI{t}{h}=\sqrt{G\WI{\rho}{c}}.\label{t_h}
\end{equation}

The unit of total angular momentum of a star can be
defined as follows:
\begin{equation}
J_0=M_0r_0^2\omega_0=4\pi\sqrt{G}r_0^5\WI{\rho}{c}^{3/2}.\label{J0}
\end{equation}

Furthermore, we need a unit of measurement for the
quadrupole moment of the system:
\begin{equation}
Q_0=M_0r_0^2=4\pi r_0^5\WI{\rho}{c}.\label{Q0}
\end{equation}

We now turn to the various contributions to the
energy of the star. In order of magnitude, gravitational
energy is equal to
\begin{equation}
\WI{E}{G}\simeq-\frac{G M_0^2}{r_0}=-G(4\pi)^2r_0^5\WI{\rho}{c}^2,
\end{equation}
kinetic energy of rotation
\begin{equation}
\WI{E}{rot}\simeq M_0(\omega_0 r_0)^2=4\pi G r_0^5\WI{\rho}{c}^2.
\end{equation}
The internal energy is represented by the integral over
the mass of the star:
\begin{equation}
\WI{E}{int}=\int\mathcal{E}dm,
\end{equation}
where $\mathcal{E}=K\rho^{\gamma{-}1}/(\gamma{-}1)$ is the specific internal
energy. Taking $n=1$ ($\gamma=2$) and expressing $K$
through $r_0$ from (\ref{r0}), we obtain:
\begin{equation}
\WI{E}{int}\simeq K\rho_0 M_0=8\pi^2G r_0^5\WI{\rho}{c}^2.
\end{equation}
As can be seen, all these energies are the same up to a
constant. Therefore, we take as a unit of energy $E_0$ the
following expression:
\begin{equation}
\WI{E}{0}=4\pi G r_0^5\WI{\rho}{c}^2.
\end{equation}

To conclude this section, we present a few more
useful relationships. The virial theorem is \cite{Tassul}:
\begin{equation}
2\WI{E}{rot}+\WI{E}{G}+3(\gamma{-}1)\WI{E}{int}=0.\label{virial}
\end{equation}
Fragmentation parameter
\begin{equation}
\tau=\frac{\WI{E}{rot}}{|\WI{E}{G}|},\label{tau}
\end{equation}
makes it possible to write (\ref{virial}) as:
\begin{equation}
\WI{E}{G}(1{-}2\tau)+3(\gamma{-}1)\WI{E}{int}=0,\label{virial2}
\end{equation}
that, given the inequalities of $\WI{E}{G}\leq 0$ and $\WI{E}{int}\geq 0$,
leads to the condition of $\tau\leq 1/2$. Finally, expressing $\WI{E}{int}$ from (\ref{virial2}), it is possible to obtain the formula for the
total energy of the star, $\WI{E}{tot}=\WI{E}{rot}+\WI{E}{G}+\WI{E}{int}$:
\begin{equation}
3\WI{E}{tot}=\frac{\WI{E}{G}\left[3\gamma(1{-}\tau){+}5\tau{-}4\right]}{(\gamma{-}1)}=\WI{E}{G}(2{-}\tau),
\end{equation}
where the last equality is given for $\gamma=2$.

\section{IDENTIFICATION OF TWINS}
We now turn to the selection of those initial stellar
configurations that are of our primary interest in this
study.
\begin{figure*}[htb]
\includegraphics[width=1.\textwidth]{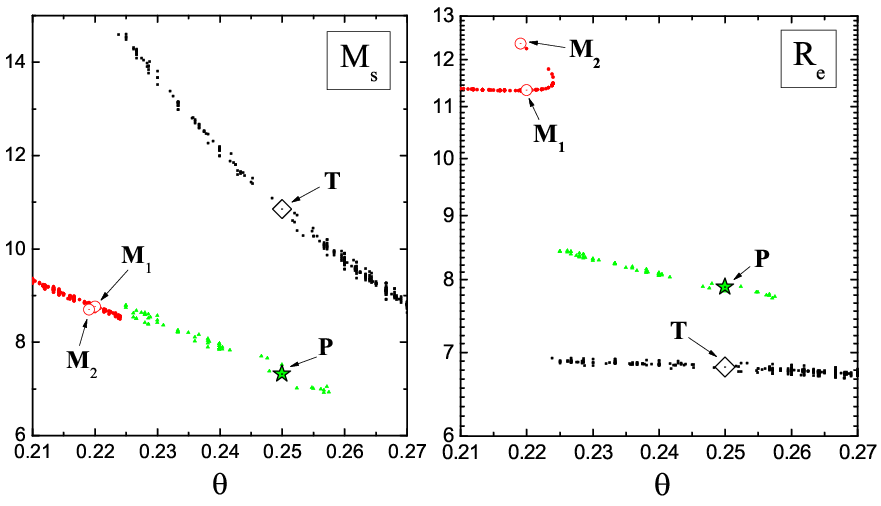}
\caption{Mass of a star, $\WI{M}{s}$ (left) and its equatorial radius $\WI{R}{e}$ (right) as functions of parameter $\theta$. The icons with indexes indicate
the configurations discussed below.}
\label{Pix:Main}
\end{figure*}
In Fig.~\ref{Pix:Main}, small sections of the diagrams $\WI{M}{s}(\theta)$ and $\WI{R}{e}(\theta)$
that are calculated in \cite{Razinkova2024} are shown. Here $\WI{M}{s}$ is
the mass of the star in $M_0$ units, $\WI{R}{e}$ is equatorial radius
in $r_0$ units (we recall that $\theta=\WI{R}{p}/\WI{R}{e}$). All configurations
have the same density at the center\footnote{In the original study \cite{Razinkova2024}, all configurations had the same maximal
density. In highly deformed stars, the density maximum
may be shifted relative to the center.} and the same
law of rotation (\ref{rotlaw}) with $\sigma=2$.

Three various types of configurations are shown in
different colors: type \textbf{T} (tori, see also~\cite{Hachisu1982}) in black, \textbf{P}
(washers) in green, and \textbf{M} (matryoshka dolls) in red;
for a detailed discussion, see \cite{Razinkova2024}. The small spread of
parameters within each group (curve of the corresponding
type) is due to the fact that the presented
data are a compilation of a large number of calculations
on various grids, some of which are sufficiently
coarse.

\begin{figure}
\includegraphics[width=0.5\textwidth]{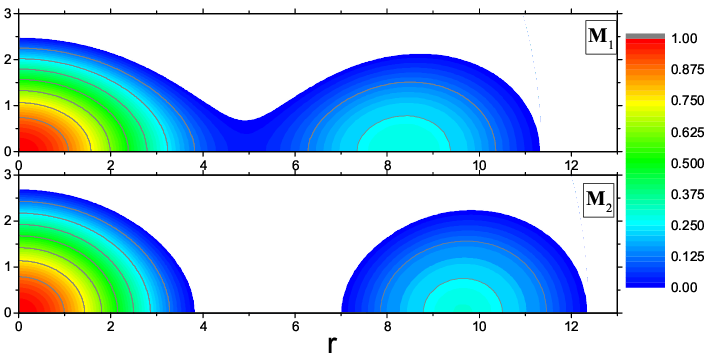}
\caption{Structure of configurations $\mathbf{M}_1$ and $\mathbf{M}_2$ that are shown in Fig.~\ref{Pix:Main}. The color indicates the density level lines.}
\label{Pix:Matryoshka}
\end{figure}

For illustration purposes, structures of two \textbf{M} configurations
of the type M are shown in Fig.~\ref{Pix:Matryoshka}.
Due to axial and equatorial symmetries, only a
quarter of the star is shown; the color indicates the
density level lines normalized to the central value.
$\mathbf{M}_1$ is a standard ``matryoshka'' with a jumper, while $\mathbf{M}_2$
is a central body surrounded by a torus (see also~\cite{Basillais2019}).

\subsection{Comparison of Twins}
We now move on to the main thing: two configurations
of types \textbf{T} and \textbf{P} that are shown in Fig.~\ref{Pix:Main} by a diamond
and an asterisk, respectively. As can be seen,
they have the same parameter value of $\theta=0.25$, but
different radii and masses (assuming the same density
at the center). To compare them, we use the results of
Section \ref{Sec:Dimensionless} and rescale the configurations to the same
mass value, $\WI{M}{s}$, by appropriately selecting the central
density values.

\begin{figure}
\includegraphics[width=0.5\textwidth]{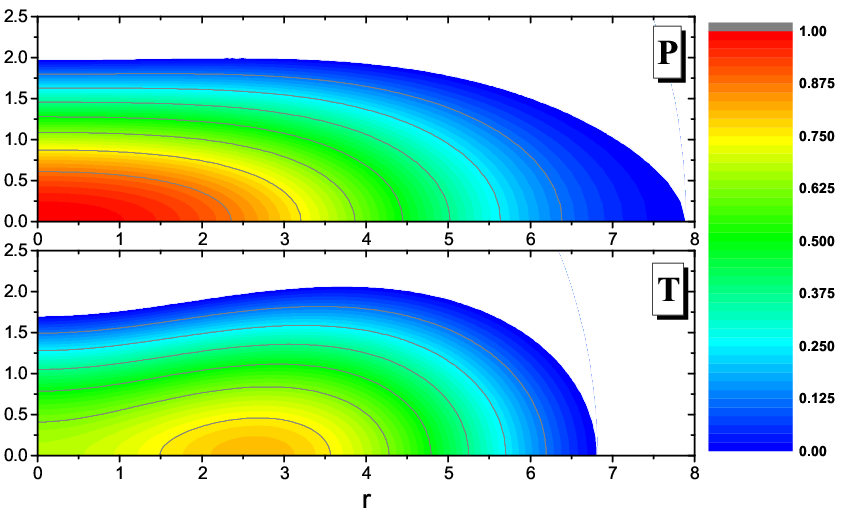}
\caption{Structure of configurations \textbf{T} and \textbf{P} that are shown in Fig.~\ref{Pix:Main}.}
\label{Pix:two_main}
\end{figure}

The structures of stars are shown
in Fig.~\ref{Pix:two_main}.
Here, unlike Fig.~\ref{Pix:Matryoshka}, the same density scale is used
for both shown configurations: the density at the center,
\textbf{P}, is taken as the unity, while the density at the
center, \textbf{T}, is selected so that the masses of the stars
coincide. As can be seen, the configurations differ very
significantly from each other, in particular, for a star of
the type \textbf{T}, unlike \textbf{P}, the maximal density lies not in its
center, but at $r\approx 2.7$. However, in all other respects,
these configurations are twins, and Table~\ref{comparison} will help to
demonstrate this.
\begin{table}
\caption{Configuration parameters \textbf{T} and \textbf{P} at the same
density at the center.}
\label{comparison}
\begin{tabular}{c|c|c}
\hline
 Parameter & Type \textbf{T} & Type \textbf{P} \\
\hline
$\WI{M}{s}$ & 10.86 & 7.32\\
\hline
$J/J_0$ &\  302.55\ &\ 166.97\ \\
\hline
$\WI{E}{tot}/E_0$ &\ -341.35\ & -158.4\\
\hline
$\tau$ & 0.2769 & 0.2764\\
\hline
\end{tabular}
\end{table}
We first turn to the last row of the table. As can be
seen, the values of parameter $\tau$ coincide with an accuracy
of up to three decimal places! We now consider
other lines. As follows from the results of Section~\ref{Sec:Dimensionless},
the quantities given here scale with density as $M\sim\WI{\rho}{c}$, $J\sim\WI{\rho}{c}^{3/2}$, and $E\sim\WI{\rho}{c}^2$. Then, as can be seen from
Table~\ref{comparison}, $M(\mathbf{T})/M(\mathbf{P})\approx 1.483$, $\left[J(\mathbf{T})/J(\mathbf{P})\right]^{2/3}\approx 1.486$
and $\sqrt{\WI{E}{tot}(\mathbf{T})/\WI{E}{tot}(\mathbf{P})}\approx 1.468$, where the configuration type is indicated in brackets. The closeness of
these numbers means that, at the same mass, these
stars will also have very close values of angular
momentum and total energy. Since their parameter $\tau$
practically coincides, then, according to the virial theorem
(see Section \ref{Sec:Dimensionless} above), their gravitational, internal,
and kinetic energies coincide with the same accuracy.
We recall that the law of rotation is the same for
them. However, these configurations themselves
undoubtedly differ from each other (see Fig.~\ref{Pix:two_main}). This is
the phenomenon of doubles.

\section{FRAGMENTATION}
A significant number of studies, both analytical \cite{Tassoul1968,OstrBoden1973}
and numerical \cite{Durisen1980,Gingold1978,Gingold1979}, are devoted to the
problem of the stability of stars and their possible fragmentation
into parts during strong rotation. Classical
results \cite{Tassul} suggest that the determining factor for fragmentation
is the parameter $\tau$ (\ref{tau}): at $\tau\gtrsim 0.14$ , secular
instability develops in the star \cite{Centrella2001,Durisen1981}, while at
$\tau\gtrsim 0.27$, instability growing on a time scale closer to
the dynamic one develops \cite{OstrBoden1973,Toman1998}. Instability basically
has the character of the so-called bar mode (see
below), although variations are possible: spirals, rings
(tori), etc.~\cite{Machida2008}. The problem of fragmentation and the
formation of a bar mode of instability was also studied
within the framework of general theory of relativity
(see, for example, ~\cite{Baiotti2007,Shibata2000}). It was found that all the
results of Newtonian theory remain qualitatively true,
and the effects of general theory of relativity only lead
to some decrease in the critical value of the $\tau$ parameter
up to the values of $\tau\simeq 0.24\div 0.25$. On the other
hand, in relativistic stars, the development of CFS
instability (Chandrasekhar–Friedman–Schutz) \cite{Chandra1970,Friedman1975,Friedman1978} caused by the action of gravitational radiation and
specific to GTR is possible. Its study is of particular
interest.

In general, the fragmentation problem has been
studied in various contexts: in the study of protostars  \cite{Bonnell1994,Bonnell1994a,Lucy1977}
 or the collapse of massive stellar cores \cite{Imshennik2010,Imamura2001,Eriguchi1985,Shibata2005}; in the accretion of matter onto a white dwarf~\cite{Yoon2004}
or the formation of binary black hole systems \cite{Zink2006,Loeb2016},
and even for the problem of the formation of the Moon \cite{Binder1978,Marov2019}.

 Below are the results of our 3D calculations of the
``twins'' dynamics and a study of their stability. The
proximity of their $\tau$ parameters(see Table~\ref{comparison}) to a critical
value of $\tau\approx 0.27$ makes these calculations particularly
interesting

\subsection{Technical Description of 3D Calculations}
All 3D numerical calculations were performed
using \textsf{FLASH}\footnote{\href{https://flash.rochester.edu/}{https://flash.rochester.edu/}}, a modern multiphysics code for astrophysical
modeling \cite{FLASH}. This is a powerful open-source software package designed specifically for research in
astrophysics and high-density and high-energy physics.
\textsf{FLASH} is a modular system that has undergone
rigorous testing. The main capabilities of the code
include the integration of various physics modules:
hydrodynamics, self-gravity, thermal diffusion, and
nucleosynthesis. \textsf{FLASH} enables the use of adaptive
mesh refinement (AMR), which allows dynamically
improving the computational accuracy in areas of
interest without significantly increasing the computational
cost. Parallel data processing makes \textsf{FLASH} an
ideal tool for high-performance computing on supercomputers,
which significantly expands the capabilities
for modeling complex astrophysical systems. The
code is constantly evolving and improving.

Most of the calculations were carried out on a computational
grid with a size of cells $100{\times} 100{\times} 100$ using
a regular personal computer. For models of particular
interest, the calculation was repeated on a more
detailed computational grid with a size of
cells $400{\times} 400{\times} 400$ using a high-performance cluster
(a comparison of calculations for these two resolutions
can be found on the author’s website \url{https://dau.itep.ru/yudin/2025/M30_LR-vs-HR.mp4}). The computing
complex is equipped with nodes based on Intel
Xeon Haswell (v3) processors, 128 GB of RAM per
node, the ability to run parallel computing on 64 processors,
and a high-performance Lustre file system.

All calculations were performed on Cartesian
homogeneous grids with the same set of parameters of
the numerical scheme. The calculation of the gravitational
potential was performed using the multipole
method for solving the Poisson equation with a maximal
value of the multipole moment equal to six. To
solve the hydrodynamic equations, a third-order precision
method (PPM) was used on smooth flows with
a limited interpolation slope. The LLF (Local Lax–Friedrichs) method was used as a solver for the Riemann
discontinuity decay problem. A free-flow
boundary condition (zero gradients) was imposed on
the entire outer surface of the computational domain.

The simulation results were thoroughly validated
by comparing data obtained on different computing
platforms and grids of different sizes. Visualization of
some of the obtained results is presented in the additional
video materials.

\subsection{Dynamics of \textbf{T}-Type Configuration}

\begin{figure*}[htb]
\includegraphics[width=1.\textwidth]{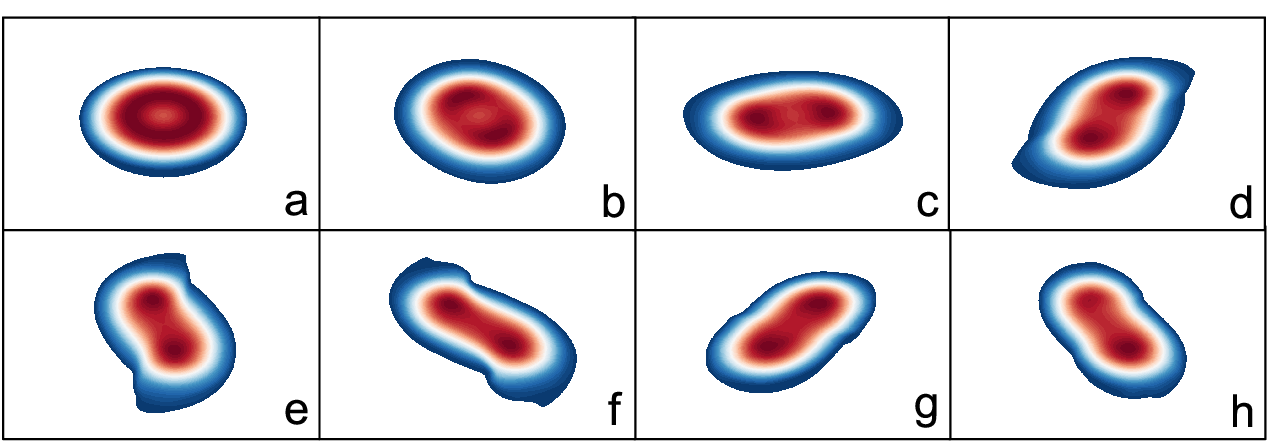}
\caption{Fragmentation dynamics of the structure of type \textbf{T} at the following times: $\WI{t}{a}=0.0$, $\WI{t}{b}=35.4$, $\WI{t}{c}=37.6$, $\WI{t}{d}=39.4$,
$\WI{t}{e}=40.8$, $\WI{t}{f}=41.7$, $\WI{t}{g}=49.7$, $\WI{t}{h}=71.2$}
\label{Pix:fragmentation}
\end{figure*}

We begin to describe the dynamics of a toroidal
configuration \textbf{T}. Eight frames illustrating the entire
calculation are shown in Fig.~\ref{Pix:fragmentation} with the density value
indicated in color. The corresponding calculation
video can be found on the author’s website \url{https://dau.itep.ru/yudin/2025/M30_zCut.mp4} (see also \url{https://dau.itep.ru/yudin/2025/M30_xCut.mp4} where the same dynamics are shown, but with a cut in the plane
perpendicular to the equatorial plane.)

Time values (in units of hydrodynamic time $\WI{t}{h}$) are
given both in the figure caption and in Table~\ref{times}.
\begin{table}[htb]
\caption{The values of the time moments shown in Fig.~\ref{Pix:fragmentation}
and~\ref{Pix:non_fragment}.}
\label{times}
\begin{tabular}{|c|c|c|}
\hline
Index & Type \textbf{T}& Type \textbf{P}\\
\hline
a & 0.0 & 0.0\\
\hline
b & 35.4 & 83.3\\
\hline
c & 37.6 & 100.1\\
\hline
d & 39.4 & 104.9\\
\hline
e & 40.8 & 107.5\\
\hline
f & 41.7 & 118.3\\
\hline
g & 49.7 & 121.0\\
\hline
h & 71.2 & 133.1\\
\hline
\end{tabular}
\end{table}

Initially, the configuration remains in a virtually
stationary state for tens of revolutions. Then, by $t\simeq 30$, disturbances begin to increase, leading to a
deviation of the star’s shape from axial symmetry
(moment b). Then, for a short time (c--e), the socalled
bar-mode of instability increases: the star is
stretched and turns into a quasi-cylindrical body rotating
around an axis perpendicular to the main axis of
symmetry. This rotation is then practically stationary.
Furthermore, two seals are visible inside the ``bar''.
Taking into account dissipative processes can make
the fragmentation complete: these two nuclei will rake
up and absorb the surrounding matter, and the initial
configuration \textbf{T} will turn into a binary system.

\subsection{Dynamics of \textbf{P}-Type Configuration}
\begin{figure*}[htb]
\includegraphics[width=1.\textwidth]{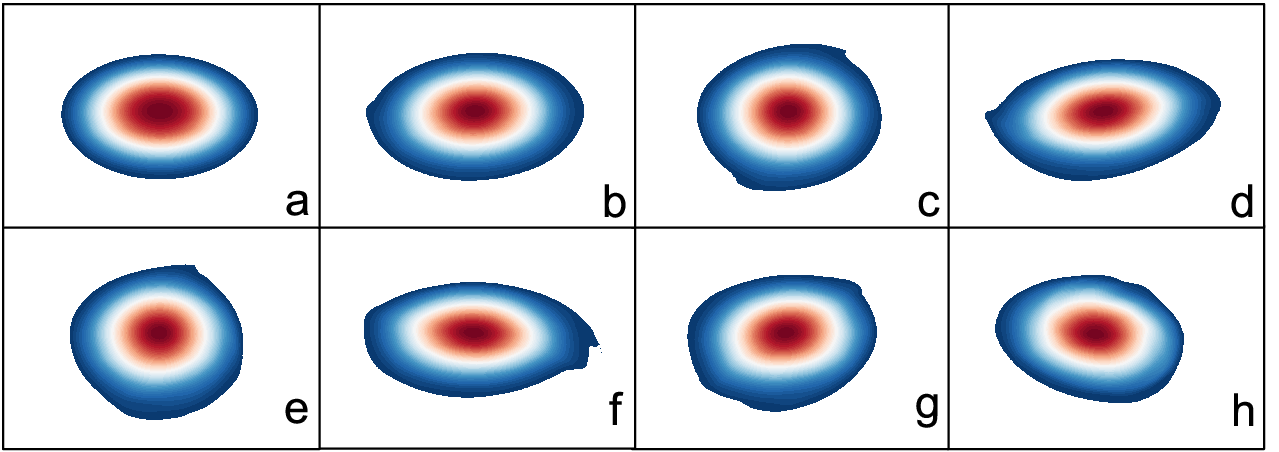}
\caption{Динамика для структуры типа \textbf{P}.  $\WI{t}{a}=0.0$, $\WI{t}{b}=83.3$, $\WI{t}{c}=100.1$, $\WI{t}{d}=104.9$,
$\WI{t}{e}=107.5$, $\WI{t}{f}=118.3$, $\WI{t}{g}=121.0$, $\WI{t}{h}=133.1$}
\label{Pix:non_fragment}
\end{figure*}
We now move on to the description of the dynamics
of configuration \textbf{P} (see Fig.~\ref{Pix:non_fragment}, as well as the calculation video \url{https://dau.itep.ru/yudin/2025/M25_zCut.mp4} and \url{https://dau.itep.ru/yudin/2025/M25_xCut.mp4}). It remains in quasi-stationary rotation
much longer than the type \textbf{T}, to $t\sim 80$. Then, the
instability of the bar-mode also begins to develop in it,
but it is much less pronounced (c--e). Furthermore,
small spiral ``sleeves'' extend from the ends of the bar.
Somewhere in the time interval around (f--g), these
spirals break away and fly away from the calculated
region, carrying away a small part (less than a percent)
of the star’s angular momentum. However, this is sufficient
for the bar mode to disappear practically and
the configuration to continue rotating, experiencing
chaotic oscillations (see also the bottom panel of Fig.~\ref{Pix:Qxy}
with the behavior of the quadrupole moments in this
case). For the interaction of the bar with spiral arms,
see, for example~\cite{Imamura2000,Williams1987}.

We stop the calculation at the moment h: by this
time, the configuration has accreted from the surrounding
background ``atmosphere''~5\%
of its own
mass and approximately the same fraction of angular
momentum, and further calculations are pointless.
Due to the escape of matter from the calculated
region, the star acquires a small momentum and
begins to ``creep'' as a whole to the left (see frame h).

To summarize, we can say that at the numerical
precision that is available to us, a configuration of the
type \textbf{P}, in striking contrast to \textbf{T}, does not show signs of
fragmentation.

Given the limitations of \textsf{FLASH}-type grid codes,
which in particular lead to the above-mentioned problems
with non-zero density of the background ``atmosphere'',
it would be interesting to repeat these calculations
using some SPH code. The SPH (smoothed particles
hydrodynamic) method is known to be free
from this drawback. The authors plan to implement
such a calculation using the well-known \textsf{PHANTOM}
code~\cite{Price2018} after eliminating some of its own problems
(see ~\cite{Potashov2024} for more details), which have already been
at least partially resolved in its latest version. As an
example of such a comparison of codes of different
types, which gave similar results, we can mention the
study~\cite{Durisen1986}.

\section{GRAVITATIONAL WAVE LOSSES}
Our calculation deals with a polytrope of $n=1$,
i.e., with an idealized model of a real star. However, it
is well known that a polytrope of index one describes
some properties of neutron stars sufficiently well, in
particular, the fact that their radius weakly depends on
their mass ~\cite{Hensel2007}. It is, therefore, interesting to estimate
whether the fragmentation process described above
could be hampered by the emission of gravitational
waves, a factor that is generally negligible for normal
stars, but extremely important for neutron or protoneutron
stars~\cite{Villain2004}. This question is also important in
the context of the rotational mechanism of supernova
explosions~\cite{Imshennik2010}: its essential ingredient is the fragmentation
of the rapidly rotating collapsing core of the star
into two (in the simplest case) parts.

\subsection{Equation for Angular Momentum}
To solve the problem described above, we write the
equation for the loss of angular momentum by the system
due to the emission of gravitational waves~\cite{Thorn_book}:
\begin{equation}
\frac{dJ^{i}}{dt}=-\frac{2 G}{5 c^5}e^{ijk}Q_{jm}^{(2)}Q_{km}^{(3)},\label{dJdt}
\end{equation}
where $e^{ijk}$ is the symbol of Levi-Civita, and $Q_{ij}$ is the
quadrupole moment of the system:
\begin{equation}
Q_{ij}=\int\!\big(r_i r_j {-} \frac{1}{3}\delta_{ij}r^2\big)\rho d^{3}r,\label{Qij}
\end{equation}
while the number in brackets above in $Q_{ij}$ expression
(\ref{dJdt}) denotes the order of differentiation (second
or third in this case). Integration in (\ref{Qij}) occurs over
the entire volume of the considered system.

We move on to dimensionless variables, assuming
$J=J_0\mathcal{J}$, $Q=Q_0\mathcal{Q}$ and $t=\WI{t}{h}\mathfrak{t}$. Then, equation (\ref{dJdt})
will take the form:
 \begin{equation}
\frac{d\mathcal{J}^{i}}{d\mathfrak{t}}=-\WI{\Lambda}{G}e^{ijk}\mathcal{Q}_{jm}^{(2)}\mathcal{Q}_{km}^{(3)},\label{dJdt_dim}
\end{equation}
where all derivatives are now taken with respect to
dimensionless time $\mathfrak{t}$.

Dimensionless parameter $\WI{\Lambda}{G}$ in (\ref{dJdt_dim}) there is:
\begin{equation}
\WI{\Lambda}{G}{=}\frac{8\pi}{5}\!\left[\frac{G\rho_0 r_0^2}{c^2}\right]^{\!\frac{5}{2}}\!\!\approx 7.5{\times} 10^{{-}5}\!\left[\frac{M_0}{M_\odot}\frac{10\mbox{km}}{r_0}\right]^{\!\frac{5}{2}}\!\!,\label{LambdaG}
\end{equation}
where we used expression (\ref{M0}) for transfer from $\rho_0$
to $M_0$.

\subsection{Calculation Results}
\begin{figure}
\includegraphics[width=0.5\textwidth]{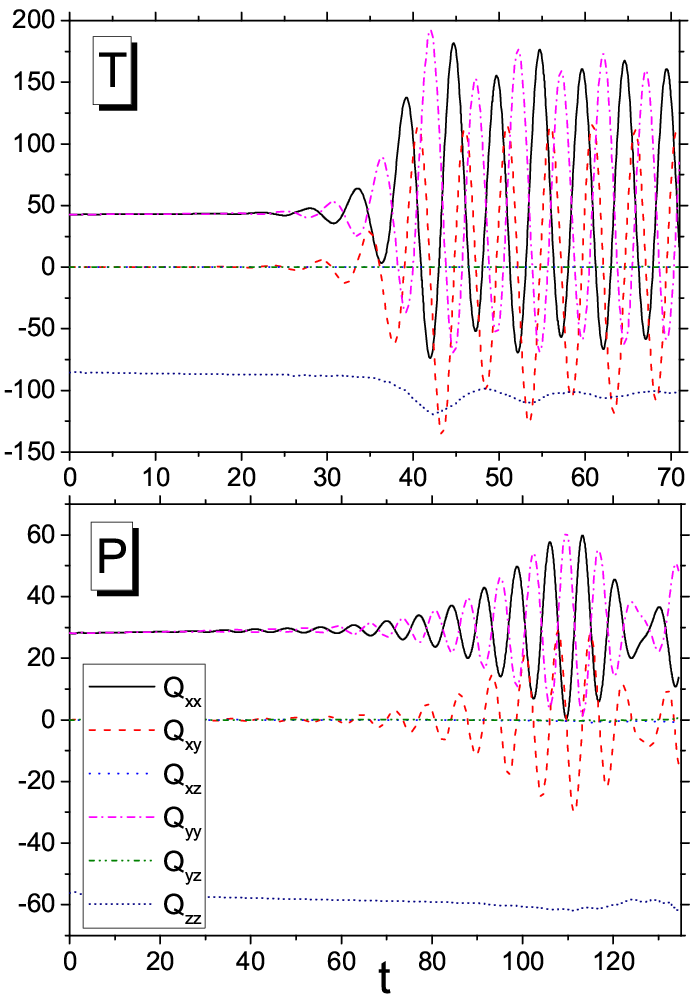}
\caption{Dimensionless quadrupole moments $\mathcal{Q}_{ij}$ as functions
of dimensionless time $\mathfrak{t}$ for configurations \textbf{T} (top)
and \textbf{P} (bottom).}
\label{Pix:Qxy}
\end{figure}

Behavior of dimensionless quadrupole moments $\mathcal{Q}_{ij}(\mathfrak{t})$
when calculating the fragmentation of a configuration
of the type \textbf{T} is shown in the top panel of
Fig.\ref{Pix:Qxy}. The $z$-axis coincides with the axis of rotation of
the star.

As can be seen, approximately until the time of
$\mathfrak{t}\sim 30$, the moments remain almost constant, then
their oscillations quickly increase and reach stationary,
almost periodic oscillations corresponding to the
uniform rotation of the resulting quasi-cylindrical
configuration (bar).

For comparison, the quadrupole moments for the
type \textbf{P} are shown on the bottom panel of the same figure.
As can be seen, this star is almost twice as long as
\textbf{T} remains in a state of uniform rotation. The scale of
subsequent oscillations is also significantly smaller,
and after the ``breakaway'' of the sleeves, these oscillations
completely die out.

\begin{figure}
\includegraphics[width=0.5\textwidth]{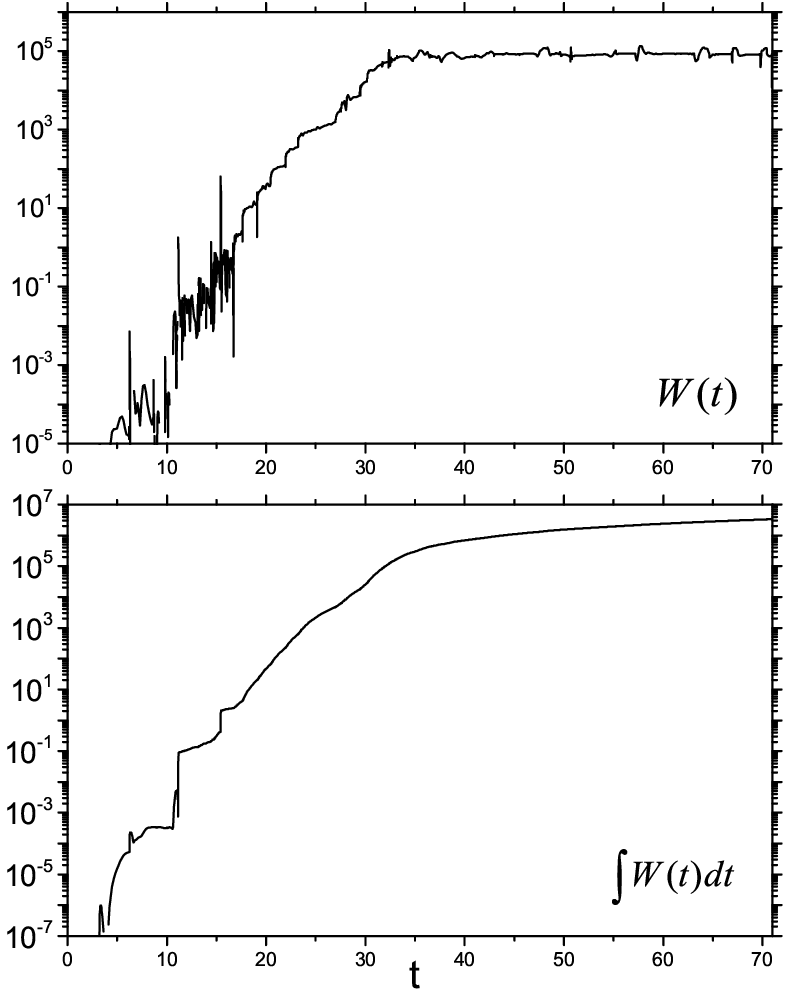}
\caption{GW losses $W(\mathfrak{t})$ (top) and their integral (bottom)
for configuration \textbf{T}.}
\label{Pix:WintW}
\end{figure}
Using the quadrupole moment values for the fragmenting
configuration \textbf{T}, we calculated the loss function:
\begin{equation}
W(\mathfrak{t})\equiv e^{zjk}\mathcal{Q}_{jm}^{(2)}\mathcal{Q}_{km}^{(3)}.
\end{equation}
To eliminate numerical noise, we first had to smooth
the calculated values of $\mathcal{Q}_{ij}(\mathfrak{t})$ with cubic splines. The
result is shown in the top panel of Fig.~\ref{Pix:WintW}. As can be
seen, until the moment of $\mathfrak{t}\sim 30$ (roughly corresponding
to panel b in Fig.~\ref{Pix:fragmentation}), the losses increase
exponentially and then reach a constant value. In fact,
fragmentation ends somewhere in the time range of
 $\mathfrak{t}\sim 30\div 35$, and after that, the configuration transfers
to a state of uniform rotation.

In the bottom panel of Fig.~\ref{Pix:WintW}, the time integral of
the loss function: according to equation (\ref{dJdt_dim})), this
value determines the total angular momentum of the
body lost to GW radiation is shown. In the time range
of $30\div 35$, this value increases from approximately $3\times 10^{4}$
to $3\times 10^{5}$. Using expressions (\ref{dJdt_dim}) and (\ref{LambdaG}), we
see that by the end of fragmentation, the carried away
angular momentum is:
\begin{equation}
\triangle\mathcal{J}\simeq (2\div 20)\!\left[\frac{M_0}{M_\odot}\frac{10\mbox{km}}{r_0}\right]^{\frac{5}{2}}.
\end{equation}
This value should be compared with the value of the
dimensionless moment of configuration \textbf{T} (see
Table~\ref{comparison}), i.e., from $302.55$. As can be seen, the losses
by the end of fragmentation do not exceed 10\% 
of the
angular momentum value and, apparently, cannot
interfere with it. However, it would be interesting to
calculate this process for various values $\WI{\Lambda}{G}$ and determine
the final fate of the fragmenting configuration
depending on the value of this parameter. Obviously,
for this, it is necessary to take into account the reverse
effect of GW radiation on the star directly in the process
of dynamic calculation. This can be done using
the formalism proposed in~\cite{Damour1990} (an example of its
implementation is given in \cite{Oohara1990}). Another possibility is
discussed in \cite{Potashov2023}. Furthermore, such an accounting of
GW radiation will make it possible to study the development
of CFS instability \cite{Kruger2021,Andersson2003}, which is of considerable interest in its own right. The authors plan to
carry out such a calculation in the near future.

\subsection{GW Signal from the Fragmentation Process}
\begin{figure}
\includegraphics[width=0.5\textwidth]{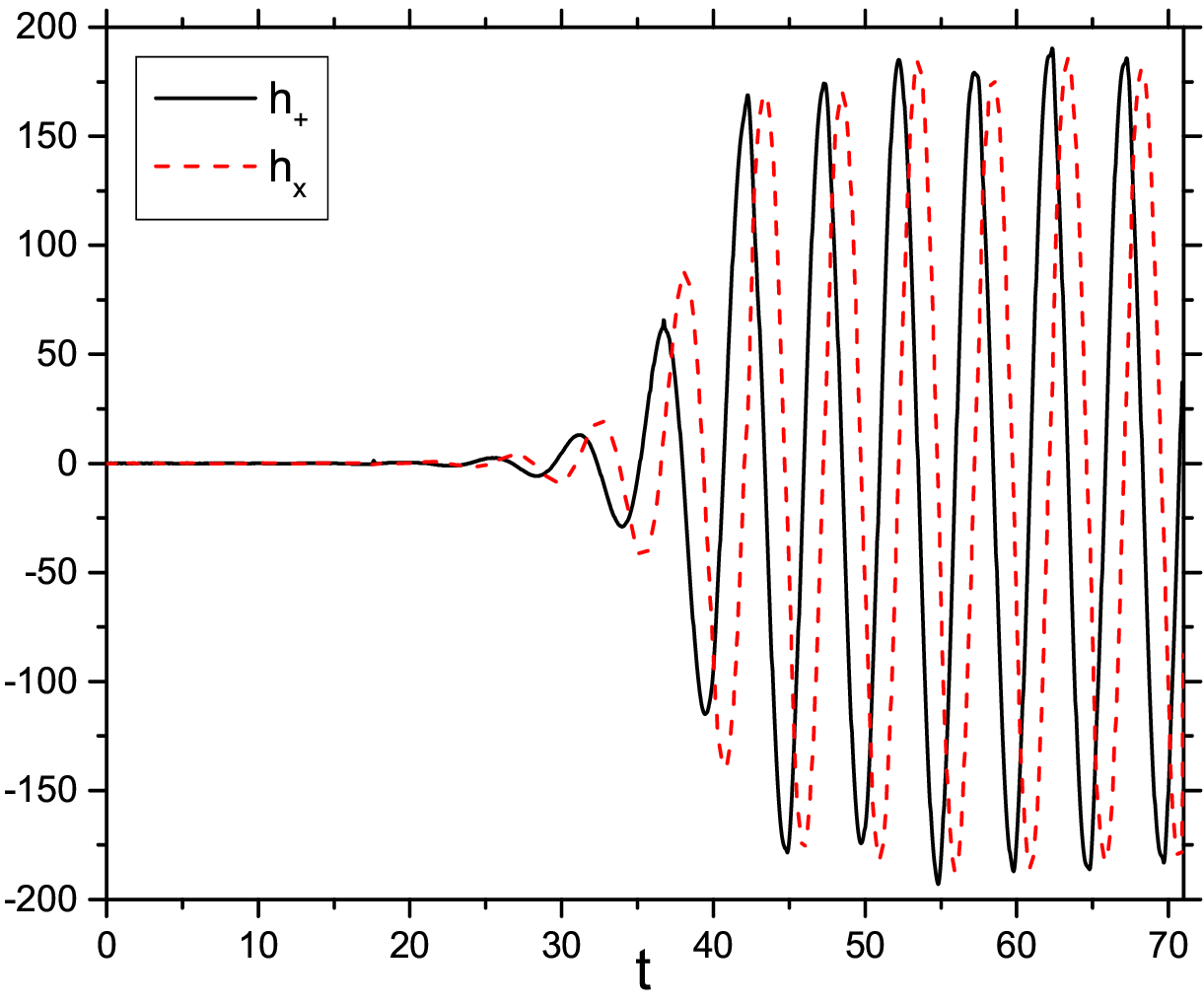}
\caption{Dimensionless fluctuations of the metric $h_{+}$ and $h_{\times}$
for configuration \textbf{T}, arb. units.}
\label{Pix:h_x}
\end{figure}

An additional result of the fragmentation process
we calculated is the possibility of obtaining an accompanying
gravitational wave signal. The dimensionless
fluctuations of the metric are~\cite{Thorn_book}:
\begin{equation}
h_{ij}=\frac{2G}{c^4 D}\frac{\partial^2 Q_{ij}}{\partial t^2}=8\pi\frac{r_0}{D}\left[\frac{G\rho_0 r_0^2}{c^2}\right]^{2}\frac{\partial^2 \mathcal{Q}_{ij}}{\partial \mathfrak{t}^2},
\end{equation}
where $D$ is the distance from the source to the receiver.
Two independent polarizations of radiation are
expressed through $h_{ij}$ as $h_{+}=(h_{xx}{-}h_{yy})/2$ and
 $h_\times=h_{xy}$. The calculation result is shown in Fig.~\ref{Pix:h_x}.

As can be seen, after the formation of the bar, the signal is
almost strictly periodic with an approximately constant
amplitude, i.e., similar to the signal from a uniformly
rotating cylinder~\cite{Thorn_book}. These results can be compared
with calculations of the GW signal from the
development of bar instability in rapidly rotating stars~\cite{Houser1998,Brown2000,Imamura2003}.

\section{DISCUSSION AND CONCLUSION}
Our study is a natural extension of ~\cite{Razinkova2024}, in which we
discovered a large number of ``exotic'' stellar configurations
with rapid differential rotation. Some of these
stellar structures were degenerate in the compression
parameter $\theta$: different configurations corresponded to
the same $\theta$ value. However, it turned out that, for
some structures, the degeneracy is even more significant:
they have practically identical integral characteristics, but differ in their internal structure. This is the
phenomenon of twins.

The next stage of our activity initially arose from
the desire to verify independently the correctness of
the \textsf{ROTAT} code that we use to calculate axial–symmetrical
configurations with stationary rotation. The
known \textsf{FLASH} code was selected for testing. The
3D modeling carried out with its help demonstrated
the stability of the considered configurations over
many tens of revolutions and thus confirmed their
dynamic stability and, therefore, the correct operation
of the \textsf{ROTAT} code and the reality of the stellar structures
we discovered.

A natural next step was a dynamic calculation at
large times for two ``doubles'' with values of fragmentation
parameter $\tau$ that are close to critical. And this
presented a surprise: despite the identity or, at least,
closeness of all integral characteristics, it turned out
that, on a long time scale, the twins behave differently.
If the configuration type \textbf{T} experiences classical
bar mode instability with complete loss of axial symmetry,
then the type \textbf{P} remains in a virtually undisturbed
state, despite the influence of the surrounding
background matter (see the video of the calculations
on the author’s page~\href{https://dau.itep.ru/sn2/team/andrey-yudin/dinamika-konfiguracii}{https://dau.itep.ru/sn2/team/andrey-yudin/dinamika-konfiguracii}).

This calculation raises the question of the real criterion
for the onset of fragmentation and the apparent
insufficiency of the condition of $\tau\gtrsim 0.27$. In \cite{Hachisu1987,Hachisu1988},
it was found that fragmentation depends both on the
topology of the rotating star and on the distribution of
angular momentum in it. In particular, the following
was found as a criterion for fragmentation of tori:
$\tau\gtrsim 0.14$.

In \cite{Saijo2006}, the existence of several types of instability:
spiral instability with one arm, and low-arm bar-mode
instability with low $\tau\sim 0.1$ and with high $\tau\sim 0.27$
was shown. In \cite{Shibata2002}, it was found that, for strongly differential
rotation, the critical value of the parameter $\tau$
may decrease to $\tau\sim O(0.01)$. In subsequent study \cite{Shibata2003}, a weak sensitivity of this result to the value of the
polytropic index $n$ was found. Finally, in \cite{Christodoulou1995}, it was
proposed to change the very definition of the fragmentation
parameter $\tau$ (\ref{tau}), and instead of the rotational
energy, use the quantity associated with the angular
momentum of the configuration. Here, it can be noted
that, firstly, it is not entirely clear how to apply this criterion
in the general case that is different from the simple
models that the authors considered, and secondly,
that the angular momentum of our ``doubles'' is practically
equal, but their behavior with respect to fragmentation
is fundamentally different. To sum up, it
must be acknowledged that this problem is still far
from being solved. We hope that the example of
twins we have considered can shed some light on
this issue.

In conclusion, we analyzed the gravitational–wave
effects during fragmentation of configuration ~\textbf{T},
including an estimate of angular momentum losses
and gravitational radiation characteristics. It was
found that, under moderate assumptions, GW losses
do not exceed ~10\% 
of the total angular momentum of
the star and, apparently, must not hinder the fragmentation
process. However, it would be interesting to
study this possibility with other input parameters with
allowance for the reverse reaction of GW radiation and
dissipative processes. The authors plan to conduct
such a study in the near future.

\begin{acknowledgments}
The authors are grateful to S.I. Blinnikov for numerous
useful discussions, particularly concerning CFS instability.
\end{acknowledgments}

\clearpage



\end{document}